\documentclass[conference]{IEEEtran}
\IEEEoverridecommandlockouts
\usepackage{cite}
\usepackage{algorithmic}
\usepackage{graphicx}
\usepackage{textcomp}
\usepackage{mathtools}
\usepackage{amsmath}
\usepackage{amssymb}
\usepackage{lipsum}

\usepackage{xcolor}
\usepackage[ruled,vlined]{algorithm2e}
\def\BibTeX{{\rm B\kern-.05em{\sc i\kern-.025em b}\kern-.08em
    T\kern-.1667em\lower.7ex\hbox{E}\kern-.125emX}}
\begin{document}

\title{A Deep Marginal-Contrastive Defense against Adversarial Attacks on 1D Models

}
\author{\IEEEauthorblockN{Mohammed Hassanin}
 \textit{University of New South Wales  }\\
Canberra, Australia \\
m.hassanin@student.unsw.edu.au
\and
\IEEEauthorblockN{Nour Moustafa}
\textit{University of New South Wales }\\
Canberra, Australia \\
nour.moustafa@unsw.edu.au
\and
\IEEEauthorblockN{Murat Tahtali}
\textit{University of New South Wales}\\
Canberra, Australia \\
murat.tahtali@adfa.edu.au
}

\maketitle

\begin{abstract}
Deep learning algorithms have been recently targeted by attackers due to their vulnerability. Several research studies have been conducted to address this issue and build more robust deep learning models. Non-continuous deep models are still not robust against adversarial, where most of the recent studies have focused on developing attack techniques to evade the learning process of the models. One of the main reasons behind the vulnerability of such models is that a learning classifier is unable to slightly predict perturbed samples. To address this issue, we propose a novel objective/loss function, the so-called marginal contrastive,  which enforces the features to lie under a specified margin to facilitate their prediction using deep convolutional networks (i.e., Char-CNN). Extensive experiments have been conducted on continuous cases (e.g., UNSW\_NB15 dataset) and discrete ones (i.e, eight-large-scale datasets \cite{datasets}) to prove the effectiveness of the proposed method. The results revealed that the regularization of the learning process based on the proposed loss function can improve the performance of Char-CNN.
\end{abstract}

\begin{IEEEkeywords}{Adversarial Attacks,  Contrastive Learning, Regularizing Hidden Space, defence models, 1D models}
\end{IEEEkeywords}
\section{Introduction}
\label{sec:intro}

Deep learning methods have been extensively involved in various real-world applications, such as cyber-security systems \cite{moustafa2019outlier}, robot applications \cite{affordance,survey}, and autonomous driving \cite{ackerman2017drive}. Though these methods showed high performance, they are vulnerable to adversarial attack techniques. Securing deep learning models has become a major need for various types of applications to trust their implementation. Attackers use different techniques to compromise the structure of deep learning models, for instance, illegally changing the model's parameters, model's inputs, or training processes. In particular, data poisoning attack is being used by attackers to alter legitimate data in the training stage that makes the model more vulnerable, less trustworthy, and degrades the whole system performance \cite{moustafa2019outlier,steinhardt2017certified}.

In order to mitigate the impact of these adversarial attacks, various types of defenses have been introduced in the literature  \cite{denoise_1}. Firstly, some of them proposed transforming input into another domain, which eliminates the added perturbation to improve the accuracy of the model against attack activity. Secondly, network modifications-based methods, which use network structures to be more defensive against attacks such as image restoration and image super-resolution approaches \cite{gao1702masking}. Finally, learning regularizers, which attempt to learn robust patterns through the training phase aiming to reduce adversarial effects \cite{papernot2016limitations,salman_adversarial}.

 The types of defenses have shown their ability to improve the accuracy against particular attacks developed by algorithms, not real-world attack scenarios, such as Denial of Service (DoS) and Distributed DoS (DDoS) \cite{moustafa2019outlier}. For instance, the regularization and network modification-based approaches are more vulnerable to white-box attacks.  Most of the literature in deep classifications, along with considering adversarial attacks, has concentrated on developing synthesized attacks to evade the models’ performances.  Few research studies have proposed defense strategies to prevent attacks against 1D data (1 Dimension), such as text data and network traffic \cite{textbugger,jones2020robust}. One of the defenses' types in the continuous domain (i.e., image) is regularizers, where an auxiliary term is added along with a loss function. This enables the training model to alleviate the impact of adversarial inputs \cite{salman_adversarial}.

 In this paper, we propose an objective function for deep-1D models. The proposed objective function is based on two main concepts: 1) boosting learning classifier's ability through marginal learning, 2) learning contrastive features to push other classes away from the corresponding class. A learning marginal classifier and contrastive learning have been proposed in various fields such as face verification \cite{center_loss,arcface}. However, our proposed method, for the first time in this domain, combines the marginal classifier and contrastive learning as one objective function to combat the adversarial impact on 1D data. To the best of our knowledge, no previous method proposed a regularizer that defends 1D models. The proposed method helps the training process to efficiently converge as well as learning contrastive features by forcing a sufficient margin between corresponding classes. The key contributions of this work are summarized as follows:
\begin{itemize}
    \item A new generative defensive method is proposed to mitigate the impact of adversarial attacks on 1D models.
    \item A new objective function represents learning features that are contrastive and guided by a marginal classifier.
    \item Extensive experiments have been conducted on nine datasets to demonstrate the effectiveness of the proposed method.
\end{itemize}

\section{Background and Related Work}
\subsection{Adversarial Attacks}

Deep learning methods have several vulnerabilities concerning security policies \cite{haq2017advanced}. These vulnerabilities include the exposure of model inputs, structure, training and inference phases, and/or model outputs. The adversarial deep learning paradigm has emerged to analyze the models' robustness as well as secure them against adversarial attacks. For instance, a slight perturbation change can degrade the models' performances. The intuition behind adversarial examples was introduced to tackle the models' vulnerabilities \cite{ball_segedy}. Goodfellow et al. \cite{FGSM} introduced a Fast Gradient Sign Method (FGSM) by generating a sample $\tilde{x}$ from a normal sample $x$ by linearizing the model's maximization. The main idea of FGSM is to move opposite to the direction of the gradient. A new version of FGSM, namely Basic Iterative Method (BIM), was proposed for the same purpose. BIM was integrated with FGSM to generate adversarial examples through multiple repetitions. Also, the generated sample was clipped to carry a small change\cite{BIM}.

 Based on BIM \cite{BIM}, Yinpeng et al. \cite{MIM} proposed a Momentum Iterative Method (MIM) as a variant of FGSM, whereas it generates adversarial samples using the momentum by introducing a new term, \textit{decay} $\mu$, to stabilize the gradient direction. Recently, Madry et al. \cite{PGD} noticed that BIM is a projected descent for negative loss function. As a result, Projected Gradient Descent (PGD) is proposed to tackle this issue by selecting some points around the clean input sample within the $L_\infty$. In general, PGD has shown its strength against various attacks as well as a strong attack type to be beaten. Adversarial attacks have not received much attention even though it is more challenging. Most of the attacks are based on heuristic methods to move, insert, modify, and delete tokens to appear inconsistent. Inspired by gradient-based perturbations in visual methods, Zhao et al. \cite{zhao2017generating} used Generative Adversarial Networks (GANs) to generate samples of the latent representation and then mapped it closer to the clean samples. Other works used different heuristic methods, such as shuffling, removing, and inserting words, into the original text \cite{alzantot2018generating}. A paraphrasing method was proposed to craft the text semantic automatically using back-translation techniques \cite{ribeiro2018semantically}. 

Jin et al. \cite{jin2019bert} used heuristics to get generated words from modified characters. They disregarded the context to increase the confusion magnitude in the perturbed text. Using synonyms also was proposed based on word embeddings \cite{gao2018black}. A phrase-level attack was studied to move, insert, modify on the level of phrases aiming to produce inconsistent text to the original one \cite{liang2017deep}. Gao et al. \cite{gao2018black} proposed a method to generate an adversarial example using two main steps: scoring important tokens and then modifying them slightly. Four scoring functions were proposed in the following order: Replace-1 Scoring (R1S), Temporal Head Scoring (THS), Temporal Tail Scoring (TTS), and Combined (THS + TTS). This scoring step is followed by a heuristic function (swap, insert, move, or delete) to generate an adversarial input. Though the recent works on natural language started to gain much attention, most of these methods are focusing on attacking text rather than securing the algorithms \cite{hao2020adversarial}.

Seeking to address the inherited vulnerability from traditional deep learning models, new methods have been proposed as defenses to improve the trustworthiness of these models. Several defenses have been introduced in the literature such as adversarial training \cite{advers_training_4}, network architecture modification \cite{network_2} or input transformation \cite{input_1}. Closest to our work, a family of defenses was developed to build more defensive classifiers against attacks. Mainly, these types of methods rely on increasing the margin or the area among the corresponding classes in the feature space to link the adversarial sample to the closest class \cite{papernot2017practical}.
Though these defenses proved effective in the continuous domain, we propose a marginal-contrastive loss function to defend against adversarial attacks.

\section{Proposed Method}

\subsection{Method Overview}

Training a classifier $f_\theta: X \rightarrow Y$, where $\theta$ is the learned parameters, which represents the relationship between $X$ and $Y$ in a pattern-wise format, such that it maps input samples to their corresponding outputs. The input space represents the training instances (\textit{e.g.} text), $X =\{x_1, x_2, ..., x_n\}$, where $x_i$ is an input instance and $n$ indicates the total number of these instances. The output space includes a representation of the possible classes, $Y=\{y_1, y_2, ..., y_k\}$, where $y_j \in R$ is the $j^{th}$ class in the output space and $k$ indicates the total number of classes.Because of adding adversarial examples to an input sample, two challenges arise in the learning schema. Firstly, this added perturbation to the input changes the pattern of the entire training process. Secondly, the classifier can not discriminate between normal and adversarial instances.

Furthermore, generating an adversarial attack draws a learning confusion in the latent space which maps it to a false class on the output space that degrades the model’s performance.  To address this issue, we propose a marginal-contrastive loss to enforce the classes' features to learn discriminatively. This type of learning will empty the shared space amongst the classes in order to make a decision properly. Learning contrastive features between the classes requires two main constraints to be fulfilled:  1) discrimination between the classes' features, and 2) contrast between positive classes and negative ones. Figure \ref{fig:pcl} shows these two constraints in detail.
 
\subsection{Marginal-Contrastive  Loss}
\begin{figure}[t]{
    \centering
       \includegraphics[width=0.90\columnwidth]{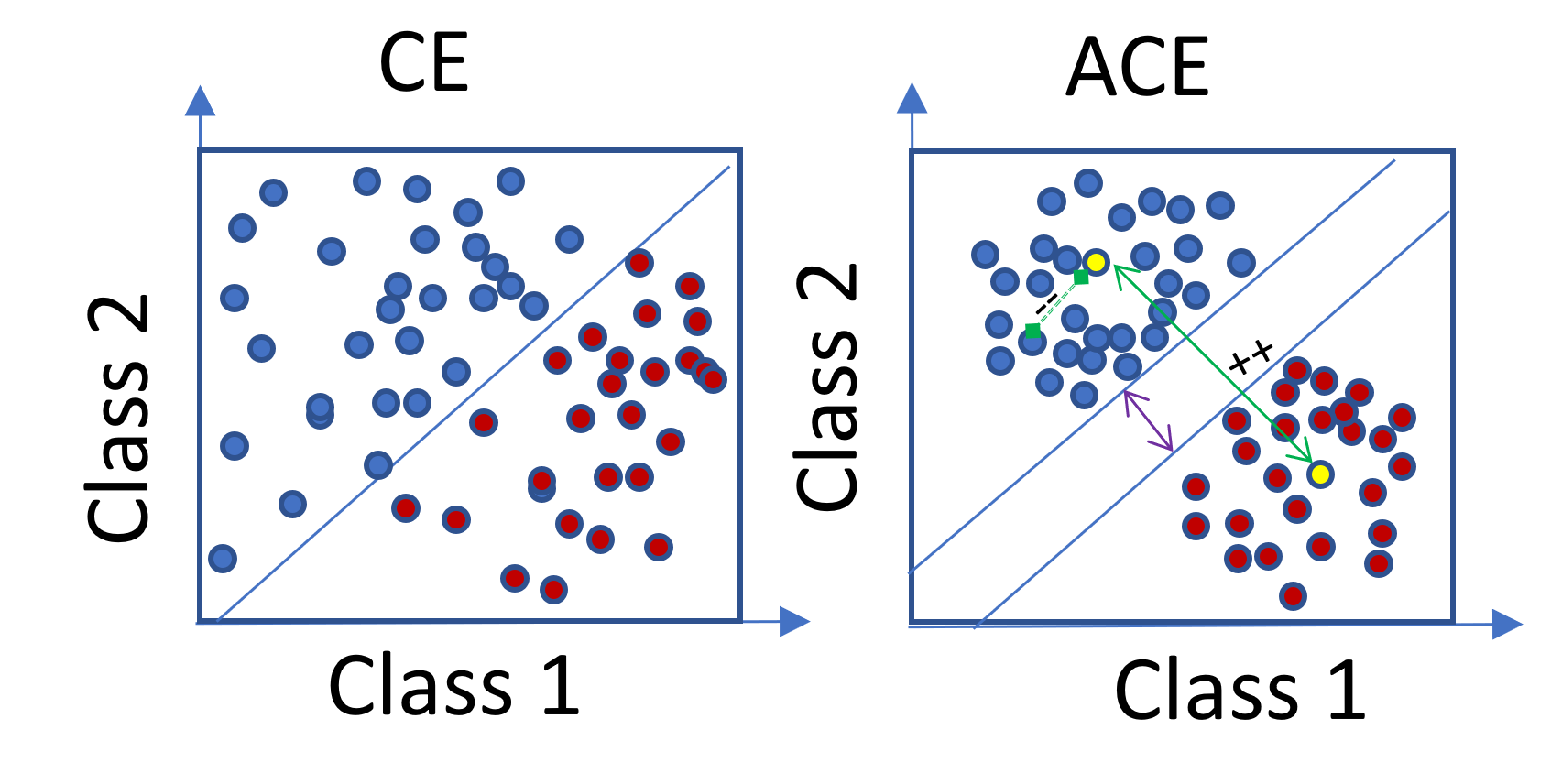}}
   \caption{A visual representation of the decision boundary between two classes using Cross Entropy loss (left) and Adversarial Cross Entropy (right). Two terms of the loss are shown as follows: 1) Marginal learning is to ensure the every-class features are assembled while emptying the decision boundary amongst classes (purple line), and 2) Contrastive to contrast between positive and negative classes (green line)}
  \label{fig:pcl} 
\end{figure}
For a long time, the main objective function for training deep learning algorithms is cross-entropy (CE) with several forms. It is responsible to map the corresponding class among the label such that each class is represented by a probability in the output space. Then, the highest class values are chosen to be the targets. The CE objective function can be formulated as follows:

\begin{equation}
\label{eq:CE}
    \mathcal{L}_{CE} = -\sum_{i}^{K}\log \frac{e^{W^T_{y_i} x_i }}{\sum_{j}{e^{W^T_{j} x_i}}},
\end{equation}

where, $\textbf{x}_i \in \mathcal{R}^d$ represents deep features of the $i$-th training sample, $y_i$ is the corresponding label. $W$ is the trainable parameters of the last linear layer, which is responsible for prediction. 

Although CE is the main loss function for most of the learning algorithms, which fails to map to the correct class in the case of boundary samples. If a sample belongs to the class $1$, but similar to the class $2$ features, CE wrongly maps it to the class $1$, as shown in Figure \ref{fig:pcl}. For instance, noisy data or poisoned data can easily trick CE. Therefore, the whole deep learning algorithms become obsolete behind CE shortcomings. Diving deeply into the origins of this problem, it was concluded that boundaries among the classes should be separated as far as possible to get rid of the overlapping between the classes' features. To address this issue, the model's features should be learned in a discrimination policy. This main idea is to force CE to learn contrastive features. This way of learning can be achieved by various methods, such as penalizing the Euclidean difference between features and their centers. This forces a margin in the latent space to widen the gap between the classes or push the classes' centers as farthest as possible from each other. The center loss proved a high improvement in the discrimination between classes' features by minimizing the intra-classes variability \cite{center_loss}. The following equation shows the center loss calculation.

\begin{equation}
\label{eq:C}
    \mathcal{L}_{dis}(\textbf{f}_i, y) = \frac{1}{n}\sum_{i} \parallel f_i - w_{y_{i}}^{c} \parallel_2^2
\end{equation}
where $(\textbf{f}_i, y)$ is the set of learned features and $w_{y_{i}}^{c} $ are set of learned weights equivalent to the size of the class. 

These weights are considered as classes' centers used to compute the factor of discrimination. The marginal softmax is used as discrimination means to predict well in the case of very high similarity amongst features, such as face verification and noisy data. 

\begin{equation}
\label{eq:marg}
    \mathcal{L}_{marg} = -\sum_{i}^{K}\log \frac{e^{W^T_{y_i} x_i - m}}{e^{W^T_{y_i} x_i - m} + \sum_{j\neq y_i}{e^{W^T_{y_j} x_i}}}
\end{equation}
where $m$ is the margin to force between the classes' features each other. 
The second constraint is to keep the distance among positive classes and negative ones as far as possible, which can be achieved by increasing the distance between the classes' centers. Marginal softmax forces the features to assemble around their centers, however, it does not ensure various features to be as far as possible. For this purpose, we need another force to balance the distribution of the features to keep the features assembled around their centers and as farthest as possible from each other.

Suppose $w^c \in \mathcal{R}^{n \times d}$, where $d$ is the size of the features while $n$ is the total number of classes. Specifically, if $w_{y_i}^c$ denotes the set of weights of the corresponding class and  $w_{j}^c$ refers to weights of other classes, the distance between $w_{y_i}^c$ and $w_{j}^c$ should be managed according to the following equation.

\begin{equation}
\label{eq:CO}
    \mathcal{L}_{contr} = \frac{1}{k}\sum_{i}\frac{\ \ \ \ \ \ \parallel x_{i} -  w^c_{y_i}\ \parallel_2}{1+\sum\limits_{\substack{j\neq i}} \parallel x_{i} -  w^c_j\parallel_2},
\end{equation}
where the numerator enforces the feature to gather around their centers and the dividend forces the negative classes' features to be away as far as possible.

The total objective function  considers two main constraints: discriminative using Eq. \ref{eq:marg}, and contrastive using Eq. \ref{eq:CO}. The supervision process will be jointly trained to ensure both discriminability conditions that are satisfied. The following equation shows the joint loss:
\begin{equation}
    \label{eq:objective}
    \mathcal{L} = \mathcal{L}_{marg} +  \mathcal{L}_{contr}
\end{equation}
\small
\begin{align}
\label{eq:total}
    \mathcal{L} =    -\sum\limits_{\substack{i}}\log \frac{e^{W^T_{y_i} x_i - m}}{e^{W^T_{y_i} x_i - m} + \sum\limits_{\substack{j\neq y_i}}{e^{W^T_{y_j} x_i}}} + 
    \frac{1}{k}\sum\limits_{\substack{i}}\frac{\ \ \ \ \ \ \parallel x_{i} -  w^c_j\ \parallel_2}{\sum\limits_{\substack{j\neq i}} \parallel w_{y_i}^c -  w^c_j\parallel_2},
\end{align}

 

\section{Experiments and Result Explanation}
In this section, the implementation details of the proposed method are summarised. Afterward, qualitative details are listed in table formats to show the results in comparison to the most recent methods.  Generally, the proposed method outperforms several state-of-the-art techniques and introduces a new defense model against adversarial attacks on selected datasets \cite{unsw15,datasets}. All the experiments were conducted on Pytorch 1.5 with Quadro 6000 24-GPU memory under Ubuntu 18.

\subsection{Implementation Details}
Char-CNN is used as the core of the proposed method as the objective function that is embedded in the architecture. We conducted our experiments on different attacks, including \cite{gao2018black}, FGSM \cite{FGSM} and PGD \cite{PGD}. The parameters of the core architecture are the default ones of Char-CNN \cite{datasets}, the number of training epochs is $100$, the learning rate is $0.001$, while the batch size is selected to be $128$.
\subsection{Results and Discussion}
\subsubsection{Continuous Attacks}
Firstly, attacks, involving FGSM and PGD, are used to hack the model.  The attacks generate various text according to the budget scale ($\epsilon$), however, it changes the text to be closer to noise than adversarial attacks. Though continuous attacks belong to the area of noise, it improves the total performance against various types of attacks. From this perspective, we compare our method to the continuous attacks to show our method effectiveness.  The proposed method is compared with three baselines as follows: 1) Cross-Entropy (CE) loss,  2) Center loss \cite{center_loss}, and 3) marginal CE \cite{arcface}.
The comparison was conducted in two cases: testing-time attacking (\textit{i.e. } 'no defense'), and adversarial training settings, which attack in the training and the testing time. For parameters, $10$ is the number of iterations for poisoning the sample, whereas the perturbation budget is $8 / 255$ and $16 / 255$, respectively. \\

\begin{table*}[ht]
\caption{Comparison between the proposed method and the baseline approaches for UNSW\_NB15 dataset.}
\centering
\begin{tabular}{|l|cccc|cccc|}
\hline
\multicolumn{9}{|c|}{\textbf{$\epsilon = 8 / 255$}}\tabularnewline
\hline
&\multicolumn{4}{|c|}{\textbf{No Defense}} & \multicolumn{4}{|c|}{\textbf{Adversarial Training}}\tabularnewline
\hline
Defenses & Char-CNN & Center \cite{center_loss}  &Marginal CE \cite{arcface}& ours & Char-CNN & Center \cite{center_loss}  &Marginal CE \cite{arcface}& ours\tabularnewline
\hline 
FGSM &49&31&51&58.5& 57&44&57.5&65.5  \tabularnewline
\hline
PGD & 40&24&41.5&52 & 47.2&34&35&58.\tabularnewline
\hline 
\multicolumn{9}{|c|}{\textbf{$\epsilon = 16 / 255$}}\tabularnewline
\hline
FGSM &41&29&38&52.9 & 40.5&36&53.8&61.2 \tabularnewline
\hline
PGD &33 &27.3&31&48& 41&31.4&27&38\tabularnewline
\hline 
\end{tabular}
\newline
 \label{table:All_mnists}
\end{table*}

For continuous attacks, we selected one dataset (UNSW\_NB15 \cite{unsw15}) to demonstrate a comparison between our proposed methods and other compelling ones. Table \ref{table:All_mnists}
 shows the comparison between our method and other methods. Table \ref{table:All_mnists} (FGSM row) for budget scale $8/255$ shows that the proposed method beats the three baseline approaches by more than $6\%$ in case of 'no defense'. Further, baselines achieved $49\%$, $31\%$, $51\%$ for CE, Center loss, Marginal CE, respectively, whereas our method was able to achieve $58.5\%$. In the case of perturbation scale  $ = 16 / 255$, our method was able to outperform all the baselines by a reasonable margin of $\approx 12\%$. PGD is known as a strong attack because it works iteratively to generate an adversarial sample rather than one step (FGSM). Table \ref{table:All_mnists} (PGD row), for budget scale $8/255$, shows that the proposed method beats the three baseline approaches by more than $10\%$ in case of 'no defense'. 

Further, the baselines achieved $40\%$, $24\%$, $41.5\%$ for CE, Center loss, Marginal CE, respectively, whereas our method was able to achieve $52\%$. Notably, Marginal CE behaves better than the other two methods with a PGD attack. In the case of perturbation scale  $ = 16 / 255$, our method was able to outperform all the baselines by a large margin of $15\%$ with PGD attack. Similarly, our method outperforms the state-of-the-art in the adversarial settings in the two various perturbation scales. Likewise, our method outperforms the baselines in adversarial training, while it achieved $65\%$ and $61.2\%$ against FGSM for the two perturbation scales.

\subsubsection{Discrete Attacks}

In typical adversarial attacks, gradients are the main contributor to guide the modifications to generate an adversarial sample. However, it is not feasible in discrete data (text-data samples) because the domain is so vast as well as it does not perceive the reader (it is a sort of noise). To this end, we use heuristic methods to generate an adversarial sample from the important tokens of the input. Then, we use swap, insertion, deletion, or move methods to craft the input and generate imperceivable text that can evade the algorithm.

Due to the lack of consensus about the methods of attacks and the less number of defenses,  results are shown with one baseline comparison (CE). Regarding databases, we use 7 publicly-used datasets \cite{datasets} along with the Spam dataset \cite{metsis2006spam} to prove the significance of our defense. This set of datasets covers various NLP-tasks such as sentiment analysis, text classification, and spam detection.  Char-CNN \cite{datasets} is used as a backbone with default parameters unless it is stated.	The perturbation budget is selected to be $30$.

\begin{figure}[!]{
    \centering
       \includegraphics[width=0.99\columnwidth]{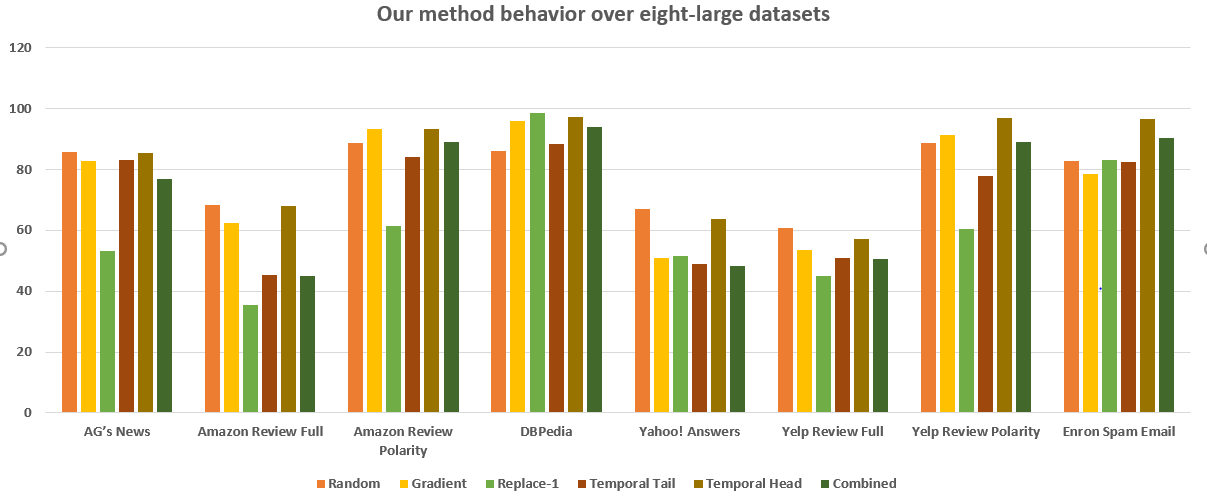}}
   \caption{A visual representation for the behavior of our method against six attacks on eight-public-large datasets \cite{datasets}.}
  \label{fig:defenceondatasets} 
  \end{figure}

Regarding generating adversarial examples, we follow \cite{gao1702masking} to attack the datasets in black-box settings. We generate 4 types of scoring combinations as follows: Replace-1 Scoring (R1S), Temporal Head Scoring (THS), Temporal Tail Scoring (TTS), and Combined (THS + TTS).  This step is followed by heuristic methods to change the input sample $x_i$. These methods can be listed as follows:
(1) Swap: Uses swapped characters to generate an adversarial sample.
(2) Substitution: Uses substituted characters to generate an adversarial sample.
(3) Deletion: Through the character's deletion, an adversarial sample is generated.
(4) Insertion:  Through the character's insertion, an adversarial sample is generated.\\
\textbf{Random \& Gradient scoring}
In this setting, a word is chosen randomly to be changed with one of the above heuristic functions. It does not have any specific method to follow. In gradient-based scoring, the gradient is used to determine the important tokens in the context. It uses the size of the gradient to decide about the important words.
In Figure \ref{fig:defenceondatasets}, our proposed method behavior is shown against eight-large datasets. Overall, the DBPedia dataset is the easiest to defend whereas securing Amazon Full Review dataset is the most difficult.

\begin{table}[ht]
\caption{Comparison between baseline CE and our method over 8-large datasets. This highlights the accuracy in the case of random scoring.}
\begin{center}
\begin{tabular}{|c|c|c|c|c|c|}
\hline 
 &Original & \multicolumn{2}{c|}{Random} & \multicolumn{2}{c|}{Gradient}\tabularnewline
 \cline{3-4} \cline{4-5} \cline{5-6}  \cline{6-6}
  & & \multicolumn{1}{c|}{CE \cite{gao2018black}} & Ours  &\multicolumn{1}{c|}{CE \cite{gao2018black}} & Ours\tabularnewline
\hline 
AG\textquoteright s News & 90.0  &89.3 & 85.6 &62.3 &82.9\tabularnewline
\hline 
Amazon Review Full & 62.0   & 61.1 & 68.3 & 47.0&62.5\tabularnewline
\hline 
Amazon Review Polarity & 95.2  & 93.9 &88.7&84.3  &93.2 \tabularnewline
\hline 
DBPedia & 98.4  &  95.2 & 86.1 & 92.9&95.9\tabularnewline
\hline 
Yahoo! Answers  & 71.0   &65.7 &67.1&43.5&51.0\tabularnewline
\hline 
Yelp Review Full  & 63.5   & 60.9 &60.8 &45.7&53.4\tabularnewline
\hline 
Yelp Review Polarity & 95.3   &95.4 &88.6&84.8&91.3\tabularnewline
\hline 
Enron Spam Email  & 95.6   &67.8 & 82.9&69.0&78.4\tabularnewline
\hline 
\end{tabular}
\end{center}
\label{table:random}
\end{table}

Table \ref{table:random} summarizes the comparison between the baseline and our method. It is clear that our method outperforms the gradient method on all the datasets with margins vary from 3.0 to 20.6 for DBPedia and AG's News, respectively. In contrast, CE baseline in random settings outperforms our method on AG's News, Amazon Review Polarity, DBPedia, Yelp Review Full, and Yelp Review Polarity dataset whereas we can beat in Amazon Review Full, Yahoo! Answers, Spam Email. Since random scoring does not follow any algorithmic way, the results are unpredictable and will be changing over the repetition.

\textbf{Replace-1 Scoring (R1S)} will be according to the following equation:
\begin{equation}
R1S (x_i) = F(x_1, x_2, ..., x_i, ..., x_n) - F(x_1, x_2, ..., \tilde{x_i}, ..., x_n)
\end{equation}
where $\tilde{x_i}$ is chosen to be out of the vocabulary to limit the search space. \\
\\

\begin{table}[ht]
\caption{Comparison between baseline CE and our method over 8-large datasets. This highlights the accuracies in the case of Replace-1 settings.}
\centering
\begin{tabular}{|c|c|c|c|c|}
\hline 
 &Original& \multicolumn{3}{c|}{Replace-1 }\tabularnewline
 \cline{3-5} \cline{4-5} \cline{5-5} 
 & & \multicolumn{1}{c||}{CE \cite{gao2018black}} & Ours& Improvement\tabularnewline
\hline 
AG\textquoteright s News & 90.0  & 30.8 & 53.2 &22.4\tabularnewline
\hline 
Amazon Review Full & 61.1   & 25.6 & 35.5&9.9\tabularnewline
\hline 
Amazon Review Polarity & 95.2  & 46.4 &61.5&15.1 \tabularnewline
\hline 
DBPedia & 98.4  & 74.9 & 98.5&23.6\tabularnewline
\hline 
Yahoo! Answers  & 71.0   & 30.0 &51.5& 21.5\tabularnewline
\hline 
Yelp Review Full  & 63.5   & 27.6 & 45.0&12.4\tabularnewline
\hline 
Yelp Review Polarity & 95.3   & 42.8 &60.5&17.7 \tabularnewline
\hline 
Enron Spam Email  & 95.6   & 76.4 & 83.0&6.6\tabularnewline
\hline 
\end{tabular}
\label{table:replace}
\end{table}

Table \ref{table:replace} shows the results of our defense and comparison with state-of-the-art.  Both methods use $\epsilon = 30$ as a distance to edit the input samples to generate adversarial samples. The results reveal that our method outperforms the baseline in Replace-1 attack by a significant margin that varies from $6.6$ to $23.6$. CE was able to get $30.8$,  $25.6$,  $46.4$,  $74.9$,  $30.0$,  $27.6$,  $42.8$, and  $76.4$, whereas our method reported $53.2$,  $35.5$,  $61.5$,  $98.5$,  $51.5$,  $45.0$,  $60.5$ and  $83.0$ for AG’s News, Amazon Review Full, Amazon Review Polarity, DBPedia, Yahoo! Answers, Yelp Review Full, Yelp Review Polarity, and Enron Spam Email, respectively.

\textbf{Temporal-Head Scores (THS)}
This type of scoring uses sequential predictions of the character $x_i$ up to $i^{th}$ subtracted from the model's prediction of $x_i$ up to $i - 1$ as follows:

\begin{equation}
THS(x_i) = F(x_1, x_2, ..., x_{i-1},x_i) - F(x_1, x_2, ..., x_{i-1})
\end{equation}

\begin{table}[ht]
\caption{Comparison between baseline CE and our method over 8-large datasets. This highlights the accuracies in the case of Temporal-head settings.}
\centering
\begin{tabular}{|c|c|c|c|c|}
\hline 
 &Original & \multicolumn{3}{c|}{Temporal Head}\tabularnewline
\cline{3-5} \cline{4-5} \cline{5-5} 
  & & \multicolumn{1}{c||}{CE \cite{gao2018black}} & Ours& Improvement\tabularnewline
\hline 
AG\textquoteright s News & 90.0  & 74.1&85.5&1.4\tabularnewline
\hline 
Amazon Review Full & 61.1   & 58.1 & 68.0&9.9\tabularnewline
\hline 
Amazon Review Polarity & 95.2  & 91.6 &93.2&1.6 \tabularnewline	
\hline 
DBPedia & 98.4  & 95.7 &97.2 &1.5\tabularnewline
\hline 
Yahoo! Answers  & 71.0   & 56.8 &63.8& 7.0\tabularnewline
\hline 
Yelp Review Full  & 63.5   & 51.3 & 57.0&5.7\tabularnewline
\hline 
Yelp Review Polarity & 95.3   & 86.5 &97.0&10.5 \tabularnewline
\hline 
Enron Spam Email  & 95.6   &85.1 &96.5 &11.4\tabularnewline
\hline 
\end{tabular}
\label{table:temporal}
\end{table}

\begin{figure}[t]{
    \centering
       \includegraphics[width=0.9\columnwidth]{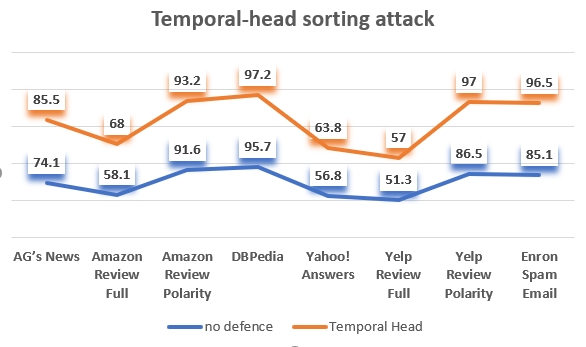}}
   \caption{Visualization of our method on the temporal-sorting-based attack over 8 datasets.}
  \label{fig:th_attack} 
\end{figure}

Table \ref{table:temporal} shows the results of our defense and comparison with the state-of-the-art.  Both methods use  $\epsilon = 30$ as a distance to edit the input samples to generate adversarial examples. The results reveal that our method outperforms the baseline in the temporal-head settings by a significant margin that varies from $1.4$ to $11.4$. CE was able to get $74.1$,  $58.1$,  $91.6$,  $95.7$,  $56.8$,  $51.3$,  $86.5$ and  $85.1$, whereas our method reported $85.5$,  $68.0$,  $93.2$,  $97.2$,  $63.8$,  $57.0$,  $97.0$, and  $96.5$ for AG’s News, Amazon Review Full, Amazon Review Polarity, DBPedia, Yahoo! Answers, Yelp Review Full, Yelp Review Polarity, and Enron Spam Email, respectively. Figure \ref{fig:th_attack} demonstrates the comparison of our method's effectiveness and CE.

\textbf{Temporal-Tail Scores (TTS)}
Temporal-tail scores are the complement of THS while it compares the tails of the sentences  as follows:

\begin{equation}
TTS(x_i) = F(x_i, x_{i+1}, ..., x_n) - F(x_{i+1}, ..., x_n)
\end{equation}

\begin{table}[ht]
\caption{Comparison between baseline CE and our method over 8-large datasets. This highlights the accuracies in the case of Temporal-tail settings.}
\centering
\begin{tabular}{|c|c|c|c|c|}
\hline 
 &Original  & \multicolumn{3}{c|}{Temporal Tail}\tabularnewline
\cline{3-5} \cline{4-5} \cline{5-5} 
 & & \multicolumn{1}{c||}{CE \cite{gao2018black}} & Ours& Improvement\tabularnewline
\hline 
AG\textquoteright s News & 90.0  &58.6&83.2&24.6\tabularnewline
\hline 
Amazon Review Full & 61.1   & 32.5&45.3&12.8\tabularnewline
\hline 
Amazon Review Polarity & 95.2&70.9&84.1& 13.2\tabularnewline
\hline 
DBPedia & 98.4  &88.2&88.3&0.1\tabularnewline
\hline 
Yahoo! Answers  & 71.0   & 35.3&49.0&13.7 \tabularnewline
\hline 
Yelp Review Full  & 63.5   &35.3& 51.0&14.7\tabularnewline
\hline 
Yelp Review Polarity & 95.3   & 71.9&77.9& 6.0\tabularnewline
\hline 
Enron Spam Email  & 95.6   & 78.7&82.4&3.7\tabularnewline
\hline 
\end{tabular}
\label{table:tail}
\end{table}

Table \ref{table:tail} shows the results of our defense and compares it with the state-of-the-art.  Both methods use  $\epsilon = 30$ as a distance to edit the input samples to generate adversarial samples. The results reveal that our method outperforms the baseline in the temporal-head attack by a significant margin varies up to $24.6$ as in AG's News. CE was able to get $58.6$,  $32.5$,  $70.9$,  $88.2$,  $35.3$,  $35.3$,  $71.9$, and  $78.7$, whereas our method reported $83.2$,  $45.3$,  $84.1$,  $88.3$,  $49.0$,  $51.0$,  $77.9$ and  $82.4$ for AG’s News, Amazon Review Full, Amazon Review Polarity, DBPedia, Yahoo! Answers, Yelp Review Full, Yelp Review Polarity, and Enron Spam Email, respectively. \\
\textbf{Combined Scores}
Combined scores mainly merging THS and TTS to benefit from both directions of importance. It can be calculated via the following equation.

\begin{equation}
CS(x_i) = THS(x_i) + \lambda\ TTS(x_i)
\end{equation}
\begin{table}[ht]
\caption{Comparison between baseline CE and our method over 8-large datasets in the case of Combined settings.}
\centering
\begin{tabular}{|c|c|c|c|c|}
\hline 
 &Original  & \multicolumn{3}{c|}{Combined}\tabularnewline
 \cline{3-5} \cline{4-5} \cline{5-5} 
  & & \multicolumn{1}{c||}{CE \cite{gao2018black}} & Ours& Improvement\tabularnewline
\hline 
AG\textquoteright s News & 90.0  & 60.4&77.0& 16.6\tabularnewline
\hline 
Amazon Review Full & 61.1   & 35.0 & 45.0&10.0\tabularnewline
\hline 
Amazon Review Polarity & 95.2  & 73.5 &89.2& 15.7\tabularnewline
\hline 
DBPedia & 98.4  &  88.8 &93.9 &5.2\tabularnewline
\hline 
Yahoo! Answers  & 71.0   &36.6 &48.3& 11.5\tabularnewline
\hline 
Yelp Review Full  & 63.5   & 38.2  & 50.5&12.3\tabularnewline
\hline 
Yelp Review Polarity & 95.3   &71.1&89.0& 18.9\tabularnewline
\hline 
Enron Spam Email  & 95.6   &75.4 & 90.2&14.8\tabularnewline
\hline 
\end{tabular}
\label{table:combined}
\end{table}

Table \ref{table:combined} shows the results of our defense and comparison with state-of-the-art.  Both methods use  $\epsilon = 30$ as a distance to edit the input samples to generate adversarial samples. The results reveal that our method outperforms the baseline in the temporal-head attack by a significant margin that varies from $5.2$ to $16.6$. CE was able to get $60.4$,  $35.0$,  $73.5$,  $88.8$,  $36.6$,  $38.2$,  $71.1$, and  $75.4$, whereas our method reported $77.0$,  $45.0$,  $89.2$,  $93.9$,  $48.3$,  $50.5$,  $89.0$, and  $90.2$ for AG’s News, Amazon Review Full, Amazon Review Polarity, DBPedia, Yahoo! Answers, Yelp Review Full, Yelp Review Polarity, and Enron Spam Email, respectively. 

\begin{figure}[t]{
    \centering
       \includegraphics[width=0.9\columnwidth]{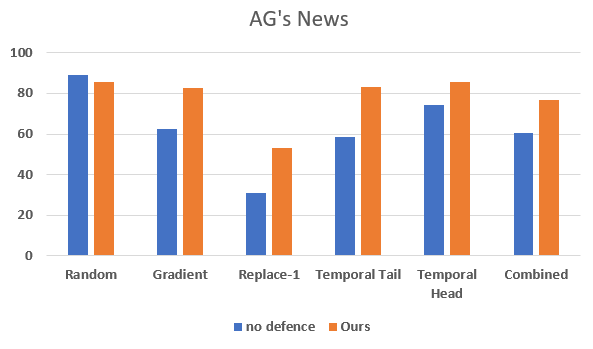}}
   \caption{A Comparison between our method and CE on AG's News datasets for various types of attacks.}
  \label{fig:ags_news} 
\end{figure}

Figure \ref{fig:ags_news} shows the significance of our method on all the datasets. It compares CE to our method. It is clear that our method outperforms in all the attacks except on random one. 

\section{Conclusions}
A new marginal-contrastive method has been proposed to defend against adversarial attacks. It is a robust method that can constitute a regularizer to defend against adversarial attacks. The experiments demonstrate that our method can address the issues of adversarial attacks on 1D models. Besides, the experimental comparisons showed that our method achieved very significant outcomes on nine-large-scale datasets under the influence of various attack settings. The impact of adversarial attacks was nearly eliminated in most of the datasets. Eventually, the proposed method highlights the need for further investigations regarding cyber-security schemes on the adversarial paradigm. In the future, marginal-contrastive learning will be extended on various architectures such as 1D, 2D and 3D models.

\bibliographystyle{plain}
\bibliography{IEEE}
\end{document}